\documentclass[reprint,amssymb,amsmath,aip,jcp,floatfix]{revtex4-1}
\usepackage{bm}%
\usepackage[dvipdfm,colorlinks=true,linkcolor=blue]{hyperref}
\expandafter\ifx\csname package@font\endcsname\relax\else
 \expandafter\expandafter
 \expandafter\usepackage
 \expandafter\expandafter
 \expandafter{\csname package@font\endcsname}%
\fi
\usepackage[dvips]{graphicx}
\usepackage{dcolumn}
\usepackage{url} 
\usepackage[latin1]{inputenc} 

\begin{document}


\title{Tunability of the optical absorption in small silver
cluster--polymer hybrid systems} 



\author{Laura Koponen$^1$}
\author{Lasse O. Tunturivuori$^1$}
\author{Martti J. Puska$^1$}
\author{Y. Hancock$^{1,2}$}
\email{yh546@york.ac.uk}
\affiliation{$^1$ Department of Applied Physics, Aalto University School of Science and Technology,\\
P.O. Box 11100, FIN-00076 AALTO,  Finland\\
$^2$ Department of Physics, The University of York\\
Heslington, York, YO10 5DD, U.K.}
\date{\today}

\begin{abstract}
We have calculated the absorption characteristics of different hybrid systems consisting
of $\textrm{Ag}$, $\textrm{Ag}_2$ or $\textrm{Ag}_3$ atomic clusters and
poly(methacrylic acid) (PMAA) using the time-dependent density-functional
theory. The polymer is found to have
an extensive structural-dependency on the spectral patterns of
the hybrid systems relative to the bare clusters. The absorption
spectrum can be `tuned' to the visible range for hybrid
systems with an odd number of electrons per silver cluster, whereas for hybrid systems
comprising an even number of electrons, the leading absorption edge can
be shifted up to $\sim4.5$~eV.
The results give theoretical support to the experimental
observations on the absorption in the visible range in metal
cluster--polymer hybrid structures.
\end{abstract}


\maketitle 

\section{Introduction}
When metal samples are reduced from bulk to nanometer dimensions,
their optical properties become governed by surface plasmons. When the
size is further decreased to few-atom clusters, the properties are
again altered as discrete energy levels determine the electronic and
optical behavior of the cluster~\cite{schmid01, zheng07}. Applications
in this size regime have become extremely important during recent
years as advanced experimental production on this scale has become
more controlled and theoretical methods are now more easily able to
access it~\cite{greiner05, balletto05}.

Silver clusters are especially interesting but challenging
materials. Their high biocompatibility suggests great potential as
functional components in biological nanomaterials ~\cite{petty04,
slocik02} with possible applications in biosensing and optical data
processing~\cite{peyser-capadona05, gleitsmann04, peyser02,
lee03}. Silver clusters readily oxidize, which can be a problem for
some applications, and methods of increasing their stability against
oxidation are actively being sought~\cite{cai98}. From the theoretical
perspective, the challenge of these systems lies in the filled
d-electron levels of silver, which makes high requirements for
modeling their electronic properties, and for determining their
potential for application.

During the last two decades, both extensive experimental and computational
studies have been done on small silver clusters (Ag$_n$) with less
than a dozen atoms. Experimentally these clusters have been studied in
stabilizing rare-gas matrices, as free silver clusters in the gas
phase dissociate by fragmentation~\cite{harbich90, fedrigo93,
schulze04, conus06}. Small silver clusters exhibit interesting luminescence and
especially fluorescence properties that can be tuned by their size and
chemical environment~\cite{peyser01, felix01, zheng02}. Interesting,
but not very well understood, experimental observations on the
tunability of their optical properties as a function of the chemical
environment have been reported. D\'iez \emph{et al}., for example,
have observed absorption in the visible range at $\sim$500 nm (2.5~eV)
from small silver clusters ($n = 1-4$) stabilized by coils
of poly(methacrylic acid) (PMAA) in the liquid phase
\cite{diez09}. The photoabsorption spectrum of this system was found
to redshift slightly when the polarity of the solvent
decreased, and other interesting properties, such as strong fluorescence
and electrochemiluminescence, were also observed. A similar synthesis
was also carried out by Shang \emph{et al.}~\cite{shang08} who assumed
that the measured absorption bands at 430~nm (2.9~eV) and 500~nm
(2.5~eV) are due to Ag$_n^+$ clusters with $n=2-8$. Their study
also reported strong fluorescence.

On the computational side, methods used to model excited-state properties
of nanosized structures include the coupled
cluster method~\cite{Cizek66}, \emph{ab initio} configuration
interaction (CI) approach~\cite{Bonacic93}, time-dependent
density-functional theory (TDDFT)~\cite{tddft-book},
Hedin's GW approach combined with the solution
of the Bethe-Salpeter equation~\cite{onida02}, and the
quantum Monte Carlo method~\cite{qmc}. The quantum chemistry
community, in particular, Bona\u{c}i\'{c}-Kouteck\'{y} \emph{et al.}
pioneered work on the absorption properties of
silver clusters using \emph{ab initio} CI, publishing many bench-mark
results~\cite{Bonacic93, bonacic99, bonacic01}.  On the DFT side,
Yabana and Bertsch were the first to use a real-time implementation of
TDDFT to calculate the optical response of small silver clusters
$\textrm{Ag}_1$, $\textrm{Ag}_2$, $\textrm{Ag}_3$, $\textrm{Ag}_8$ and
$\textrm{Ag}_9^+$ in the gas phase~\cite{yabana99}. A more systematic
real-time TDDFT study of the optical properties of $\textrm{Ag}_n$ with
$n = 1-8$ followed~\cite{idrobo05}. Both of
these studies showed good agreement with experimental results with
a rather rigid difference of about 10\% in the excitation energies.
More recent TDDFT studies have been performed on
larger $\textrm{Ag}_n$ clusters with $n\ge 4$
~\cite{tiago09, zhao06, harb08}.

A hybrid structure consisting of a silver cluster and an organic
compound is of great interest as this is already a reliable model for
a silver cluster that has been functionalized by its
environment. Examples of studies on silver-cluster--organic-compound hybrid
systems include silver-carboxylates for which DFT methods have been used
to determine the shapes of the resulting silver
nanoparticles~\cite{olson06, kilin08}. The interaction with the organic
compound can cause strong modifications to the optical properties
of the cluster. For example, Mitri\'c \emph{et al.} studied the absorption
spectra of silver cluster--tryptophan hybrid systems using the TDDFT~\cite{mitric07}.
They observed a strong size- and structural-dependence of the photoabsorption and photofragmentation in
these systems. Other examples of recent TDDFT work on silver-cluster--organic-compound hybrid structures include a study of the enhancement mechanism in surface enhanced Raman
scattering~\cite{morton09, zhao07, jensen07} and the determination of the
optical properties of silver-cluster--biomolecule-hybrids~\cite{tabarin08}.

In this paper, we study the effect of organic compounds on the
photoabsorption spectra of silver cluster systems. For our model systems, we choose clusters consisting of one to three silver
atoms attached to fragments of the PMAA polymer. This choice is
motivated by recent experiments in Refs.~\cite{diez09, shang08,
konradi05} where small silver clusters have been formed in the
presence of PMAA in different solutions. In our calculations we focus on the cluster-polymer system only and omit the solvent for simplicity. This approach has also been used by Mitri\'c \emph{et al.}
who calculated the excitation properties of interacting nanoparticle-biomolecule subunits~\cite{mitric07}.

The paper is organised as follows. The selected geometries for our systems are investigated and
presented in Sec.~II. The DFT and TDDFT methods are covered in Sec.~III. Sec.~IV
contains the results and the analysis of the ground-state calculations
and Sec.~V contains the results of the optical absorption
(excited-state) calculations. Finally, the conclusions are given in
Sec.~VI.

\section{Structures}

\begin{figure*}[ht]
\begin{center}
\includegraphics[width=0.91\textwidth]{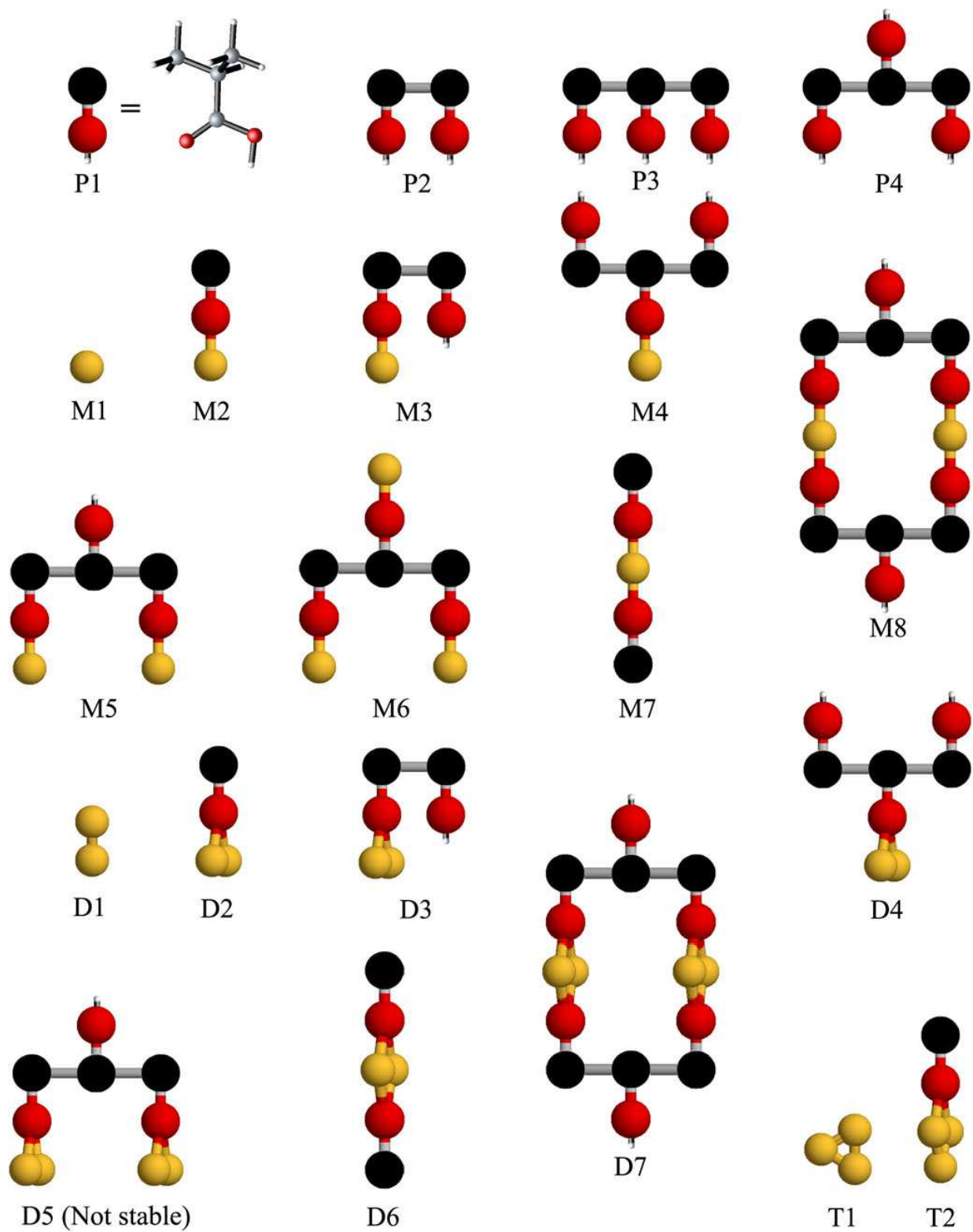}
\end{center}
\caption{\label{fig:structures}(Color) Schematic pictures of studied
structures (not to scale). Black, dark red, tiny white and small
yellow dots represent PMAA units, their carboxylate groups, the hydrogen atoms of the carboxyl groups, and Ag atoms,
respectively. The bond lengths or bond angles in the pictures do not
represent the calculated values except for the detailed illustration
of polymer-monomer P1, where large grey, large dark red and small light
grey dots represent the carbon, oxygen and hydrogen atoms, respectively.}
\end{figure*}

In order to emphasize their essential features, a schematic representation of the structures we have studied is shown in Fig.~\ref{fig:structures}. Full geometry optimization on these systems have also been performed, and the details of these calculations can be found in Sec.~III. The acronyms used for these structures consist of a descriptive letter (P = polymer, M = monomer Ag$_1$ (atom),
D = dimer Ag$_2$, T = trimer Ag$_3$) followed by a consecutive number. Our choice of systems for the hybrid structures is motivated by experimental evidence, which shows that carboxylate
groups are capable of bonding with Ag ions in polyacids (see,
e.g., Ref.~\cite{konradi05} and references therein). Hence, in this work, our primary interest has been to study small (i.e., single atom and dimer) hybrids. For the monomer
(M1--M8) and dimer (D1--D7) structures, a large variety of different
geometries have been systematically studied, whereas only a few trimer
structures (T1--T2) have been used to test whether the trends obtained
for monomer atom and dimer structures can be extended for systems containing
larger Ag$_n$ clusters.

The pure polymer (P1--P4) and pure cluster (M1, D1 and T1) geometries
are used as reference systems throughout this work. A simple,
representative structure for a polymer-cluster hybrid system has one
Ag$_n$ cluster attached to the polymer-monomer unit, for example,
R--COO$^-$Ag$_n^+$ (M2, D2 and T2). We also consider more complicated
structures where the Ag$_n$ cluster is between the carboxylate groups of
two such polymer-monomers (M7 and D6). Here, the Ag$_n$ cluster is
bound to four oxygen atoms, which create a stabilizing environment for it.

The case of longer polymer chains bonded to Ag$_1$ monomers and Ag$_2$ dimers
is also addressed. Two questions arise regarding the
spectral properties of these systems. First, does the length of
the polymer chain itself have an effect on the absorption
characteristics of the Ag$_n$ cluster (M2 vs.~M3 or M4, and D2 vs.~D3
or D4)? Second, are the clusters attached to nearby branches of the
system interacting with each other (M2--M4 vs.~M5 and M6 or M7 vs.~M8, and D6
vs.~D7)?

When considering longer polymer chains, the concept of tacticity plays
an important role in determining the lowest-energy structures. For a
pure polymer, both iso- (P3) and syndiotactic (P4) isomers were
considered. A similar comparison can be made between M3 and M4 as well
as between D3 and D4, even though the chain lengths are not equal. All
of the structures with three polymer units in a row were chosen to be
syndiotactic for two reasons. First, there is experimental evidence
to suggest that the systems are likely to be syndiotactic~\cite{diez09}. Second,
isotactic polymer chains with Ag$_n$ clusters attached to neighboring
polymer units tend to aggregate to form larger clusters. Thus, isotactic
polymer chains do not favor the formation of stable hybrid structures
with the small Ag$_n$ clusters that are studied here. In the case of Ag$_2$ clusters,
the neighboring clusters aggregate even when they are attached to the next-neighbor polymer units
for both iso- and syndiotactic polymers. The unstable structure D5 is
shown in Fig.~\ref{fig:structures} as an example to high-light this
point. In comparison, the structure D7 in which both dimers have stable
anchorings to two polymer chains is found stable.

The relative energies of the different structures were not studied in detail
as it is not possible to compare the energetics of systems having
different numbers and types of atoms. A direct energy comparison can only be
made between the structures M2 and D6, and between the structures M5
and D7. In these cases, the D6 and D7 structures are favoured in
energy by at least 1~eV per Ag atom.

\section{Computational methods}

The SIESTA program~\cite{siesta02}, which implements the
standard Kohn-Sham self-consistent density-functional method, was used in spin-dependent calculations for the geometry optimizations.
The Perdew-Burke-Ernzerhof (PBE)
generalized gradient approximation (GGA)~\cite{pbe},
Troullier-Martins pseudopotentials~\cite{tm2}, and the double zeta
plus polarization (DZP) basis-set were also used in this work.
The SIESTA program optimizes the geometries of the systems through a series of atomic displacements, until the calculated forces on the atoms are less than 0.01~eV/Å. This is the standard procedure for geometry optimization using the SIESTA code, in which the numerical ``noise'', e.g., due to
the real-space grid in the potential calculation, prevents the geometry optimization from being stuck at high-symmetry saddle points. Hence, in this work, Hessian analysis was not used for the structural optimization.

The TDDFT program OCTOPUS~\cite{octopus06} (version
3.0.0/3.0.1) was applied in all further ground-state and
excited-state calculations. Real-time TDDFT (RT-TDDFT) in its various
flavors has been successfully used to determine the spectral
properties of a large variety of both metal clusters and organic
compounds~\cite{elliottreview08}. Calculating optical absorption
spectra by TDDFT  and comparing them with the measured spectra has proven to be
an efficient tool for determining
cluster and molecule structures, and even distinguishing between
different isomers (see, e.g.,
Refs.~\cite{castroreview,c20-paperi}). In most cases, a good agreement
with experiments up to a few tenths of an eV in spectral features is achieved.

In this work, we used the RT-TDDFT method as implemented in OCTOPUS to calculate the absorption spectra in the linear response regime. The
RT-TDDFT implementation follows Ref.~\cite{yabana96}, where the system
is instantaneously excited by applying an electric field, and the
occupied Kohn-Sham orbitals are then propagated in time. The
photoabsorption cross-section follows from the induced dipole moments,
and is calculated separately for each of the three spatial
directions. The RT-TDDFT method should not be confused with
linear-response TDDFT (LR-TDDFT) methods, such as the Casida
method~\cite{casida96}. LR-TDDFT methods are superior for calculations
of small systems. However, due to their poor scaling, the RT-TDDFT
method becomes advantageous for large systems as it scales linearly
with respect to the electron number. In this study, the Casida
LR-TDDFT method within OCTOPUS was used to resolve the
specific transitions contributing to the absorption spectra for systems
M1, M2, D1, and D2.

The PBE exchange-correlation functional was used in the TDDFT
calculations, except in the Casida calculations, where the available
kernel in OCTOPUS corresponds to the local density approximation
(LDA). Troullier-Martins pseudopotentials \cite{tm2} were used in all
of the TDDFT calculations in this work. The calculations in the OCTOPUS program
are based on a real-space grid. We used a spacing of 0.23 Å for the real-space grid, and a simulation radius around each atom of 8 Å, which are typical values
in real-space RT-TDDFT calculations. The validity of the grid-spacing
was verified by reference calculations for the test structure M2 using
grid-spacings of 0.17~Å and 0.20~Å. All three grid-spacings produced
similar absorption spectra up to an accuracy of about 0.1~eV in the relevant
low-energy region of about 0--7~eV. All relevant characteristics of
the spectra were clearly resolved using this accuracy, as the low-energy excitations
of small metal clusters are well-spaced. The obtained numerical accuracy
is also within the general accuracy of a few tenths of an eV for this
method~\cite{tddft-book, castroreview}. The time-step for the calculation was
0.0025~$\hbar/\textrm{eV}$ and the number of time-steps was 20000 for
the pure Ag$_n$ clusters and the PMAA-monomer structure and 15000 for the
rest of the structures, resulting in a total propagation time of
$50~\hbar/\textrm{eV}\approx 33$~fs / $37.5~\hbar/\textrm{eV}\approx
25$~fs. In the Casida calculations, the number of unoccupied states that were
taken into account was 4.2 times the number of occupied
states for the largest structure D2 and at least 8 times the number of
occupied states for the other structures. These numbers were considered adequate
to produce reliable results for the D2 structure and high-quality
results for all of the other systems.

\section{Ground-state results}

The calculated bond lengths for the Ag containing structures
are shown in Table~\ref{table:bonds}. In the case when the dimer is attached to
the polymer, the Ag-Ag bonds are slightly lengthened (D1 vs.~D2--D7). For the trimer, the Ag-Ag bond lengths in
the triangle become almost equal (T1 vs.~T2). The Ag-O bond lengths
vary only slightly between 2.17--2.42~Å in the different
structures, which is somewhat longer than the 2.04~Å bond length in
bulk silver oxide Ag$_2$O~\cite{wyckoff63}.

\begin{table}[tbp]
\caption{\label{table:bonds}Properties of the studied
systems. Even (odd) denotes even (odd) number of electrons
per Ag$_n$ cluster. All bond lengths are given in
Ångströms. Cluster-cluster distances are in italics.}
\begin{center}
\begin{tabular}{lllll}
\hline Structure & Even/Odd & Ag-O & Ag-Ag           \\
\hline
M1 & Odd & --         & --             \\
M2 & Even      & 2.32--2.34 & -- \\
M3 & Even      & 2.32--2.42 & --             \\
M4 & Even      & 2.33--2.34 & --             \\
M5 & Even      & 2.33--2.34 & -- \\
M6 & Even      & 2.32--2.35 & --             \\
M7 & Odd       & 2.25--2.27 & --             \\
M8 & Odd    & 2.25--2.28 & \em{6.62} \\
\hline
D1 & Even      & --         & 2.581          \\
D2 & Odd       & 2.27--2.32 & 2.709          \\
D3 & Odd       & 2.26--2.28 & 2.710 \\
D4 & Odd       & 2.27--2.28 & 2.709          \\
D6 & Even      & 2.17 & 2.781          \\
D7 & Even      & 2.17       & 2.786, \em{6.52}\\
\hline
T1 & Odd      & --         & 2x2.644, 3.023 \\
T2 & Even      & 2.24--2.25 & 2x2.74, 2.723  \\
\hline
\end{tabular}
\end{center}
\end {table}

The ground-state calculations also give the HOMO-LUMO energy gaps,
which can be considered as zeroth-order approximations to the first
excitation energy. We found that the structures of the same type, i.e.,
those having the same cluster size and the same number of polymer chains (for
example, M2--M6 and D2--D4), have energy gaps of similar size. On
the other hand, a strong gap variation exists between structures of different
types. An odd number of electrons per Ag$_n$ cluster is found to introduce a
HOMO-LUMO energy gap smaller than 1.5~eV in all of the polymer-containing
structures, whereas all polymer-containing structures with an even number
of electrons per Ag$_n$ cluster have gaps that are larger than 1.85~eV.
Even though M8 has a total number of electrons that is even, the
odd number of electrons per Ag$_1$ monomer in this system leads to local magnetic
moments in the vicinity of the two Ag atoms, and to a HOMO-LUMO energy
gap that is smaller than 1.0~eV. For comparison, Refs.~\cite{zhao01, pereiro07}
also report on calculated HOMO-LUMO energy gaps smaller than 1~eV for odd-electron
$\textrm{Ag}_n$ clusters within the size range of $n = 2-20$. The exception to this trend is the single silver atom (M1 in our notation), which has a gap that is greater than 3~eV.

\section{Excited-state results}

\begin{figure}[htbp]
\begin{center}
\includegraphics[width=0.50\textwidth]{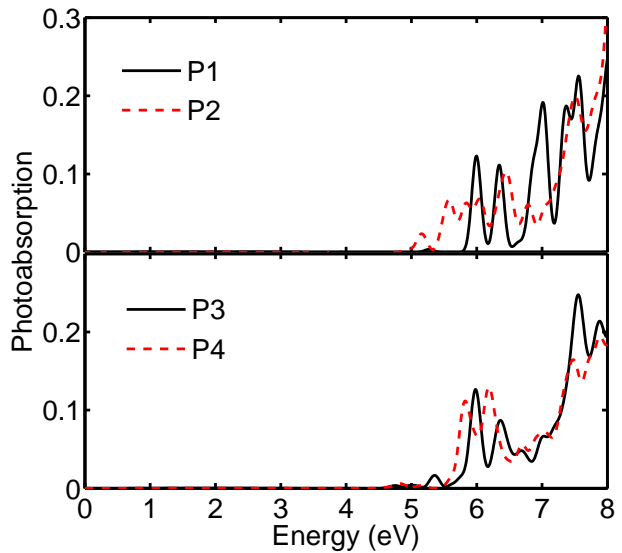}
\end{center}
\caption{\label{fig:p1-p4}(Color) Photoabsorption spectra of the pure
polymer structures. The absorption strengths are scaled down by a
factor of 1/2 for P2 and 1/3 for P3 and P4 to compensate for the
different system sizes.}
\end{figure}

The absorption spectra for the pure polymer structures are given in
Fig.~\ref{fig:p1-p4}. Each system shows a similar spectrum with an
absorption edge above 4.5~eV. In the energy range of 4.5--7~eV,
several peaks appear with slightly varying patterns depending on the
type of structure. The spectra of structures P3 and P4, which differ
only by tacticity, are almost identical in this energy range. The largest
spectral strengths for these systems appear at higher energies
ranging up to several tens of eV's. This region, which is above the ionization energies of these systems, is not included in this figure as it is beyond the range of application of the TDDFT method.

The absorption spectra corresponding to structures with monomers Ag$_1$,
dimers Ag$_2$ and trimers Ag$_3$ are shown in Figs.~3, 4 and 5,
respectively. The pure Ag$_n$
clusters, M1, D1 and T1, have their main lowest excitation energies at
3.8~eV, 3.0/4.5~eV and 2.5/2.8/3.3/3.6~eV, respectively, in
excellent agreement with previous studies~\cite{idrobo05, yabana99}.

\begin{figure}[htbp]
\begin{center}
\includegraphics[width=0.50\textwidth]{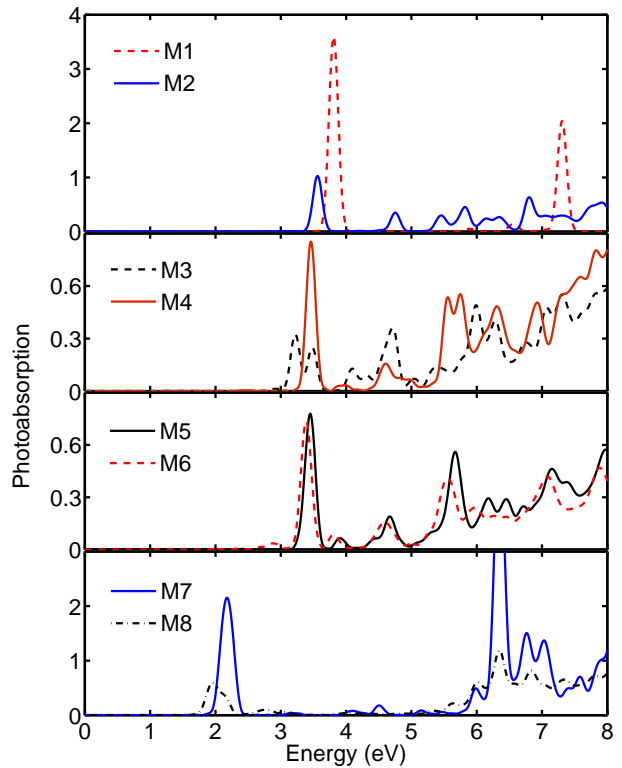}
\end{center}
\caption{\label{fig:m1-m8}(Color) Photoabsorption spectra of the
structures with Ag$_1$ monomers. The absorption strengths are scaled
down by a factor of 1/2 or 1/3 for systems containing 2 or 3 Ag
atoms, respectively, to compensate for the different system
sizes. Note the different y-axis scales.}
\end{figure}

\begin{figure}[htbp]
\begin{center}
\includegraphics[width=0.50\textwidth]{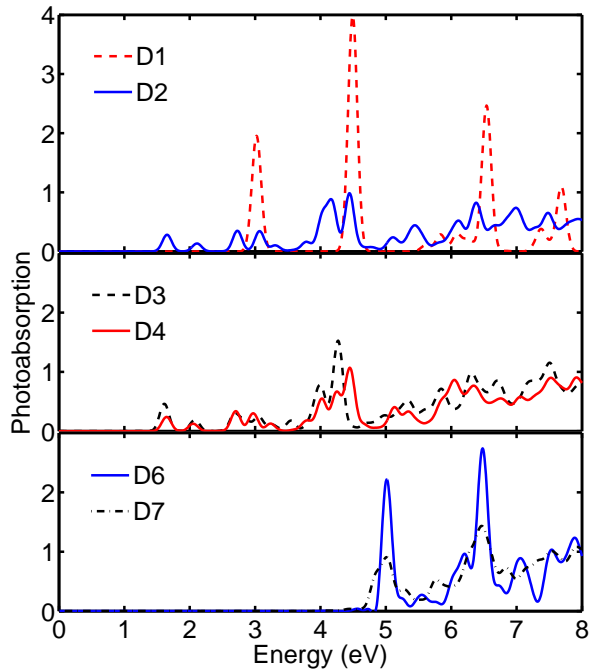}
\end{center}
\caption{\label{fig:d1-d7}(Color) Photoabsorption spectra of the
structures with Ag$_2$ dimers. The absorption strength is scaled down
by a factor of 1/2 for D7 to compensate for the different system size.}
\end{figure}

\begin{figure}[htbp]
\begin{center}
\includegraphics[width=0.50\textwidth]{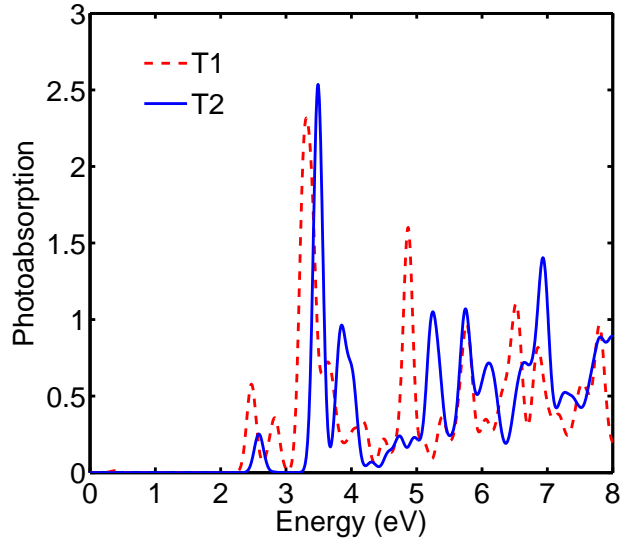}
\end{center}
\caption{\label{fig:t1+t2}(Color) Photoabsorption spectra of the
structures with Ag$_3$ trimers.}
\end{figure}

The inclusion of the PMAA monomer in structures M2 and T2 causes only
slight changes of a few tenths of an eV in the position of the lowest
excitation and a slight decrease in its intensity relative to those of the
Ag monomer M1 and trimer T1, respectively (cf.~M1 and M2 in
Fig.~3 and T1 and T2 in Fig.~5). In addition, the spectral features of
the polymer above 4.5~eV appear in both the M2 and T2 cases. In contrast,
for the dimer D2 in Fig.~4, the main low-energy excitations are
perturbed with new excitations appearing at surprisingly low energies
of 1.7~eV and 2.1~eV. From the individual directional components of the
spectrum (not shown in the figure), it can be concluded that these two
low-energy excitations are caused by charge oscillations in the dimer-axis
direction, whereas the excitations at 2.7--4.5~eV are along all three
spatial directions. To compare, the pure dimer D1 excitation at
3.0~eV stems from the dimer-axis direction, and the doubly-intense
excitation at 4.5~eV arises from the two other degenerate spatial
directions perpendicular to the dimer-axis. Hence, we conclude that
both low-energy peaks in the pure dimer spectrum have been destroyed
and are replaced by a new pattern of multiple low-energy peaks in the
spectrum of the D2 structure. We conjecture that this change in the
low-energy spectrum arises due to the odd number of electrons in
the vicinity of the Ag$_n$ cluster corresponding to
the perturbation in the electronic structure induced by
the polymer. Such a perturbation is already reflected in the ground-state
results, for example, in the widths of the HOMO-LUMO energy gaps.

Elongating the polymer chain causes only minor changes in the
spectra. For example, compared to the single polymer-monomer systems M2 and
D2, the spectra for the corresponding asymmetric structures M3 and D3
containing chains of two polymer-monomers, and the structures M4 and D4
containing chains of three polymer-monomers,
differ only slightly (see Figs.~3 and 4). Neighboring
Ag$_n$ clusters do not significantly affect the spectrum either. This
can be seen, for example, in Fig.~3 by comparing the spectrum of
the M5 structure consisting of the syndiotactic chain of three
polymer-monomers and the Ag atoms attached to next-neighbor polymer-monomers,
with that of the structure M6 with the syndiotactic chain of
three polymer-monomers and the Ag atoms attached to each of the
polymer-monomers.
Only a slight redshift of about 0.1~eV and an extra shoulder at
about 2.7--2.9~eV are observed in the latter case. To summarize,
adding polymer units or even Ag atoms in M3--M6 does not change
the fundamental trends in the spectra compared to M2. The same holds
for the dimer structures D2--D4.

When the Ag$_n$ clusters are surrounded by two PMAA chains, dramatic
changes in the absorption spectra are observed. For example, the
low-energy excitations of M7 and M8 correspond to the odd number of
electrons per $\textrm{Ag}_n$ cluster in the hybrid structure.
This is similar to the odd electron number effect seen in the D2, D3 and D4 systems, which
resulted in low-energy excitations. For the Ag monomer structure M7
(Fig.~3), the main low-energy excitation appears at
2.2~eV. Thus, this excitation is about 1.3~eV lower in energy than the
low-energy excitations in M2--M6. The structure M8, which contains two such
clusters at a mutual distance of 6.6~Å, shows a moderate splitting
and a slight redshift of the main excitation with respect to M7.

In the case of dimers that are surrounded by two polymer chains (D6
and D7, Fig.~4), an opposite effect is observed. Namely, the
absorption edge is brought up to over 4~eV. The low-energy excitation
of the pure Ag dimer D1 (Fig.~4) at about 3~eV in the dimer-axis
direction is totally suppressed in the spectra for the D6 and D7
systems. The main excitations at about 5~eV and 6.5~eV for the D6 and
D7 systems arise from the direction of the ``bridges'' between the two
polymer chains. Thus, the presence of the polymer causes quenching of
the low-energy absorption of the Ag$_2$ dimer. Because no quenching
was observed when Ag$_n$ clusters were attached to single polymer
chains (M2--M6, T2), we deduce that the formation of the symmetric
structure of the Ag$_n$ cluster and the four oxygen atoms in the
D6 and D7 systems plays a key role in the suppression of certain
directional vibration modes.

To examine the origin of the shifts in the excitation spectra, the
photoabsorption spectra of the smallest structures M1, M2, D1 and D2
were calculated using the Casida method. This enables us to determine
which orbitals are involved in the excitations, as such information is
not directly available from the RT-TDDFT calculations. The
compositions of the main low-energy excitations of M2, D1 and D2 are
shown in Figs.~6a--8a. For the pure Ag monomer M1, the transition
from the HOMO to the triply degenerate LUMO makes a 95\% contribution
to the low-energy excitation. Note that an almost constant shift of
about 0.3~eV in transition energies between Figs.~3--4 and 6a--8a is
caused by the use of the LDA kernel instead of the PBE kernel in the
Casida calculations. To demonstrate the extent of the electronic
perturbation induced by the polymer, the orbital
illustrations of a few orbitals around the gap are also shown in
Figs.~6b--8b, where the orbitals are illustrated as constant-value
surfaces of the electron wavefunctions.

\begin{figure}[htbp]
\begin{center}
\includegraphics[width=0.45\textwidth]{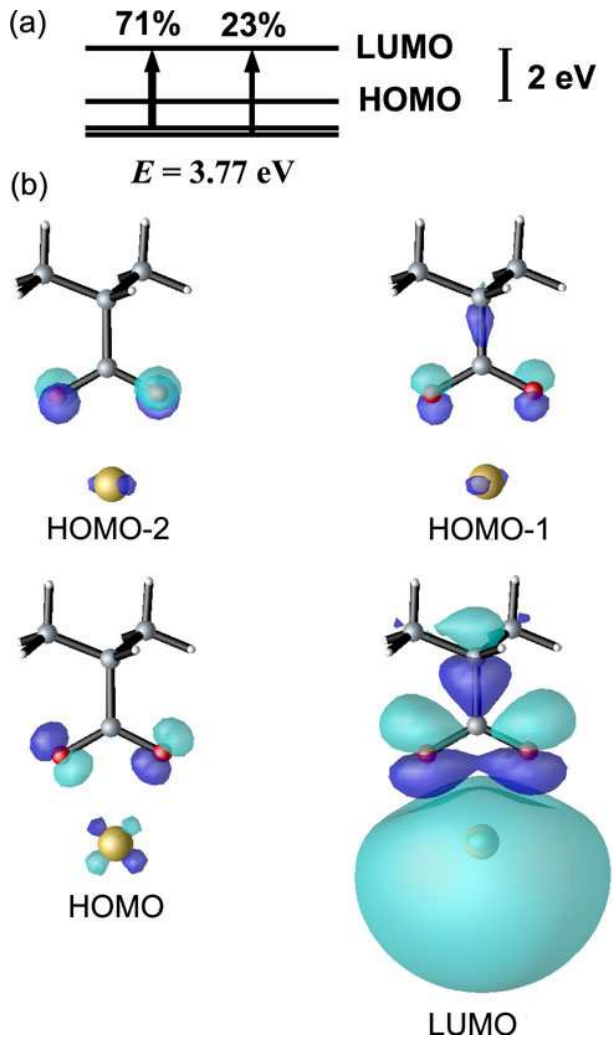}
\end{center}
\caption{(Color) (a) Energy level diagram of the main low-energy
excitation at energy $E$ and (b) orbital illustrations of the structure M2.
Light blue represents positive and dark blue negative parts of the
wavefunction.}
\label{fig:energiesM2}
\end{figure}

Fig.~6a shows that the main low-energy excitation in the absorption
spectrum of M2 corresponds to the transitions from the
HOMO-1 and HOMO-2 orbitals to the LUMO orbital. As shown in
Fig.~\ref{fig:energiesM2}b, these orbitals are localized over the
carboxylate group and the Ag atom, and even over the carbon skeleton. This
differs clearly from the pure Ag atom HOMO-LUMO excitation
from the s-orbital to the triply degenerate p-orbital.

For the Ag dimer D1, the LUMO is an anti-bonding orbital and the
degenerate LUMO+1 and LUMO+2 are $\pi$-bonding orbitals as can be seen
in Fig~7. A similar pattern with a clear interaction between the Ag
dimer and the carboxylate group is seen in the orbitals of the D2
structure in Fig.~8. The difference between D1 and D2 is that D2 has an odd number
of electrons, and a net spin resulting in an absorption spectrum with
several excitations at lower energies.
Previous results by
Mitri\'c \emph{et al.}~\cite{mitric07} for
$\textrm{Ag}_n$ cluster--tryptophan structures show similar features to our
findings. For the Ag dimer attached to the carboxylate group of a tryptophan molecule,
an excitation at an anomalously low energy of about 2.3~eV was observed~\cite{mitric07}.
This effect was associated with a HOMO-LUMO transition that is similar
to the one described above for our system, D2.

Next, we consider the quenching of the lowest excitation peak when
moving from the dimer D1 to structures D6 and D7 (See Fig.~4). The
increase in the bond length between the Ag atoms is
far too small to explain the quenching. Instead, we note that the
HOMO-LUMO transition makes a 94\% contribution (Fig.~7) to this excitation.
The quenching that is observed is therefore due to the destruction of the $\sigma$ shape of the
HOMO, which happens already in the D2 structure as seen in Fig.~8.

Finally, we want to consider the charge transfer as a possible source of error
in our calculations. The effect of charge
transfer can lead to errors of several eV's in excitation energies for
the commonly used (semi)local functionals such as PBE~\cite{peach08}. The
hybrid functionals such as B3LYP do not perform much better.
To probe the reliability of the PBE functional in our calculations, we
have used the criterion by Peach \emph{et al.} to estimate the
amount of charge transfer~\cite{peach08}. In this criterion, Peach \emph{et al.} define the spatial
overlap $\Lambda$ as attaining values between 0 and 1, so that large values
correspond to low charge transfer. According to Peach \emph{et al.}
the PBE results are predicted to be reliable for $\Lambda>0.4$.
We found that for the structure M2, $\Lambda$ is for the lowest energy
excitation within the range of 0.50--0.54. For the structure D2, the
$\Lambda$ values for the five low-energy
excitations shown in Fig.~8a are, in the order of increasing excitation
energy, within the ranges of 0.71--0.73, 0.60--0.63, 0.60--0.62, 0.43--0.46
and 0.61--0.64, respectively. The uncertainty in
these values is due to the inclusion of a limited number of orbitals in
the calculation of $\Lambda$. From these values and the orbital
illustrations in Figs.~6 and 8, we conclude that the character of the
transitions is mainly local, and that the PBE functional
can be safely used to study the main features of the excitations in
these systems. To see whether the same conclusion can be drawn for our
larger systems, we have examined the electronic structure of M5. We
observe that the orbitals near the gap of this system have shapes similar to those of
M2 (See Fig.~6). Moreover, each of the orbitals of M5 is totally or mainly
localized in the vicinity of one Ag atom only. Supported by the
previous result that the photoabsorption spectra changed only
slightly with the chain length, we can conclude that the
excitations of M5 are also local in character. Using the structure D2
as a reference, a similar analysis was also performed
for D6, which yielded the same result. As the rest of the structures have a strong
resemblance to these test cases, we do not expect the charge transfer
to be a major source of error in any of the excitations in the studied
structures.

\begin{figure}[htbp]
\begin{center}
\includegraphics[width=0.47\textwidth]{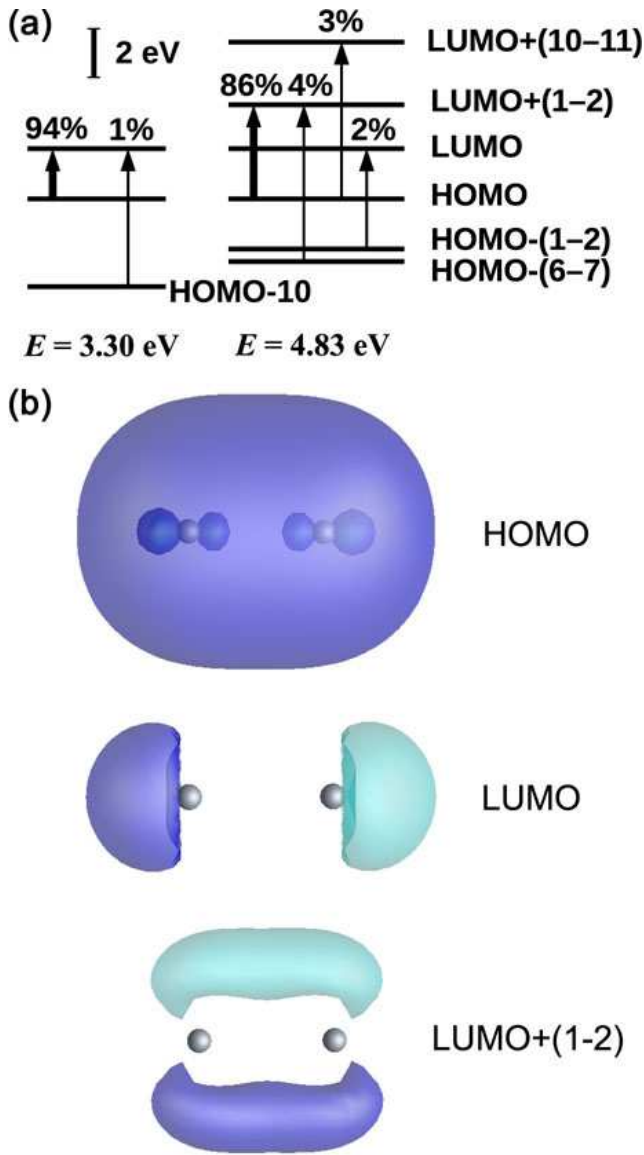}
\end{center}
\caption{(Color) (a) Energy level diagram of the main low-energy
excitations at energies $E$ and (b) orbital illustrations of Ag dimer D1.
The excitation at $E=4.83$~eV is doubly degenerate as the LUMO+1 and
LUMO+2 states are degenerate orbitals. Light blue represents positive
and dark blue negative parts of the wavefunction. Reference energy
level diagrams can be found in Refs.~\cite{idrobo05, tiago09}.}
\label{fig:energiesD1}
\end{figure}

\begin{figure}[htbp]
\begin{center}
\includegraphics[width=0.4\textwidth]{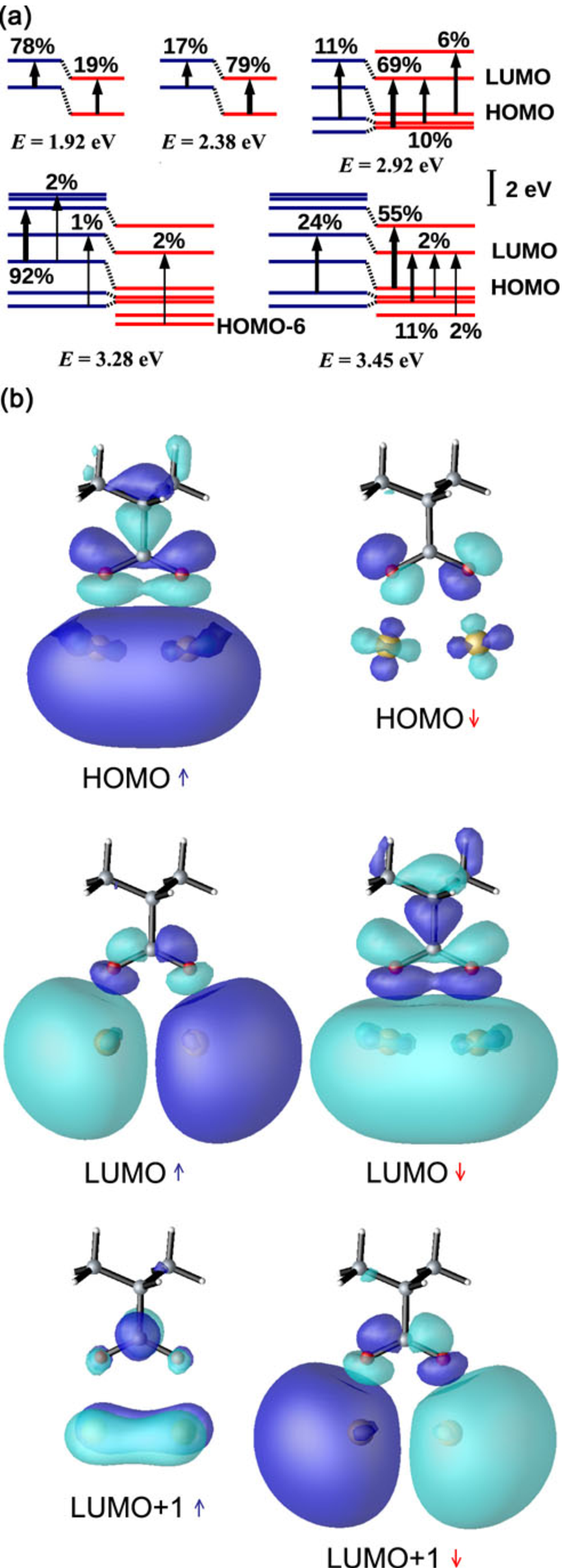}
\end{center}
\caption{(Color) (a) Energy level diagram of the main low-energy
excitations at energies $E$ and (b) orbital illustrations of
the structure D2. The left and right (blue and red) energy levels
represent the two different spin populations. Light blue marks the
 positive and dark
blue the negative parts of the wavefunction.}
\label{fig:energiesD2}
\end{figure}

\section{Conclusions}

We have studied how the photoabsorption spectra of Ag$_1$, Ag$_2$ and
Ag$_3$ clusters are affected by the presence of the PMAA polymer. The
structure optimization shows that there exists stable configurations
in which the Ag$_n$ clusters are bound to the carboxylate group of the
polymer.

We observe that the bonding of the polymer to the Ag$_n$ clusters has
an extensive effect on the absorption characteristics in certain
cases. Namely, when the number of electrons per Ag$_n$ cluster is odd,
the electronic structure and thereby the spectral features of the
Ag$_n$ clusters are strongly perturbed. For example, for the
PMAA$^-$Ag$_2^+$ systems with an odd number of electrons, the minimum
absorption energy is lowered by almost 1.5~eV resulting in absorption
in the visible range. Suppression of the lowest excitations is observed when an
Ag$_n$ cluster with an even number of electrons, such as Ag$_2$, is
surrounded by two carboxyl groups, so that the cluster is bonded to
four oxygen atoms.

Our results reveal that the absorption features of the smallest Ag$_n$
clusters are very sensitive to their chemical environment. This
suggests that these clusters are viable candidates for materials with
tailored optical properties. Already, there exists experimental
evidence on the ability to tune the absorption and emission spectra of
small Ag$_n$ cluster--polymer hybrid systems. The results presented
here give theoretical confirmation of those findings, thus showing
such systems to be promising candidates for developing functional
nanomaterials with adjustable properties.

\begin{acknowledgments}

This research is supported by the Academy of Finland through the
Centers of Excellence Program (2006--2011). CSC, the Finnish IT center
for science, is acknowledged for providing computer resources (project
tkk2035). We thank Robin Ras and Isabel D\'iez for introducing us the
experimental motivation of this study and Risto Nieminen for
discussions.

\end{acknowledgments}

\end{document}